\documentstyle[aps,prl,psfig]{revtex}
\begin{document}

\title{Effect of spin orbit scattering on the magnetic and 
superconducting properties of nearly ferromagnetic
metals: application to granular Pt}
\author{D. Fay and J. Appel}
\address{I. Institut f\"ur Theoretische Physik,
Universit\"at Hamburg, Jungiusstr. 9, 20355 Hamburg, 
Germany}
\date{\today}
\maketitle
\begin{abstract}
We calculate the effect of scattering on the static, exchange enhanced,
spin susceptibility and show that in particular spin orbit scattering  leads
to a reduction of the giant moments and spin glass freezing temperature
due to dilute magnetic impurities. The harmful spin fluctuation
contribution to the intra-grain pairing interaction is strongly reduced  
opening the way for BCS superconductivity. We are thus able to explain
the superconducting and magnetic properties recently observed in 
granular Pt as due to scattering effects in single small grains.  
\end{abstract}
\pacs{ }
\vspace{0.5in}

The recent observation of superconductivity in Pt grains of 
$\approx 1 \mu$m size at $\approx 1 $mK \cite{HermPRL} motivated this 
theoretical study of the superconducting and magnetic properties of small
grains taking account of the spin-orbit (s-o) scattering by external and 
internal surfaces\cite{Herm}. The importance of the s-o interaction at
surfaces has been shown by Merservey and Tedrow\cite{Merservey}  
from a number of different measurements on superconductors. In Pb, 
for example,the probability that a conduction electron will change its 
spin direction in a surface collision is of the order 1. Taking into account 
s-o scattering, we show that the interplay between incipient magnetism 
and superconductivity in Pt is tilted towards BCS superconductivity 
because s-o scattering is inimical to magnetism and reduces the 
paramagon effects that inhibit singlet pairing in the late transition metals.

Why can s-o scattering generate superconductivity (sc) in Pt or other 
nearly ferromagnetic metals? In bulk Pt, no sc is observed despite a strong
electron-phonon coupling; the BCS parameter is 
$\lambda^{Pt}_{Ph}\approx 0.4\,$ \cite{Pinski}. The reason for the absence of
sc in bulk Pt is due to the strong exchange interactions between the itinerant
5d electrons. The resulting paramagnon effects suppress BCS s-wave sc. The 
Pt grains are small enough to have a large surface to volume ratio but are 
sufficiently large ($\ge 100\AA$) that the Bloch representation applies and 
we can ignore the Rashba effect \cite{GorRash}. Although the 
"lattice softening" near surfaces may enhance $\lambda^{Pt}_{Ph}$,
more importantly, the s-o scattering at rough surfaces strongly reduces the 
harmful  paramagnon effects. In the case of Pt grains, the extremely weak 
impurity magnetism observed at mK temperatures clearly points to an 
important role of the changed magnetic behavior for the occurence of sc
\cite{HermPRL,HermPR}. Independently of whether the sc extends 
throughout the grain or is restricted to a surface shell of thickness small
compared with the grain size, s-o scattering at surfaces and defects will 
be important for the sc and magnetic properties. If shells exist in
the compacted granules, they may (as in a thin film) consist of
small crystallites large enough for bulk superconductivity but
sufficiently small to limit the mean free path for s-o 
scattering. We find that, with reasonable values for the exchange and 
scattering parameters,  sc in granular Pt is possible at the observed 
temperatures.
 
We first address the magnetic properties of small grains by calculating the
static susceptibility $\chi(q)$ in the presence of ordinary and s-o scattering, 
taking the exchange enhancement effects into account as in 
Ref.~\onlinecite{FayAppel}. The Stoner factor $S$ accounts for the
average exchange and the spin correlation range $\sigma$ measures the 
spatial range of the inter-atomic exchange. Scattering is included by 
considering the effect on $\chi_0\,$, the susceptibility without 
exchange enhancements. We find a significant effect of s-o scattering on 
$\chi_0$ that affects both $S$ and $\sigma$.

The susceptibility $\chi(r)$ is then calculated to determine 
how scattering as well as exchange enhancement affects the the RKKY 
(Ruderman-Kasuya-Kittel-Yoshida) oscillations occuring in $\chi_0(r)$. The
short range and long range parts of $\chi(r)$ determine the two pertinent
magnetic properties observed in dilute magnetic systems
\cite{HermJLTP1,HermJLTP2}, namely, the magnitude of the giant moment
$\mu_{gm}$ and the scale of the spin glass freezing temperature, $T_f\,/\,x$, 
in, e.g., $\mbox{PtFe}_x$. The exchange effects suppress the RKKY 
oscillations at small r yielding the ferromagnetic correlations responsible
for $\mu_{gm}\,$; s-o scattering reduces $\mu_{gm}$. The spin glass 
transition observed in the bulk $\mbox{PtFe}_x$ systen is due to the long 
range oscillations of $\chi(r)$ relevant for the interaction between two 
impurity moments at a distance $r \gg a(=\mbox{lattice constant})$ in the
absence of scattering effects. With scattering  $\chi(r)$ at large $r$ is so 
strongly reduced that spin glass freezing would not be expected in the Pt 
grains where $x=4\,$ppm\cite{HermPRL}.

We consider first the magnetic properties.
The static q-dependent, exchange enhanced susceptibility $\chi(q)$ in the 
presence of ordinary and s-o scattering is appropriate to describe the 
giant moments and spin glass freezing. We assume a susceptibility of the
RPA form:
\begin{equation}
\chi(q)=\frac{2\mu^2_B \, \chi_0(q)}
{1-  \chi_0(q) \frac{1}{3} \left[ U + 2J_H + 3J'(q) \right] } \, ,
\label{chi1}
\end{equation}
where $\chi_0(q)=N(0)\,u(q)\,$, $N(0)$ is the density of states (DOS) per spin
for all three 5d sub bands at the Fermi level, and $u(q)$ reduces to the
Lindhard function for free electrons with no impurity scattering. 
$U$ is the intra-atomic self-exchange and $J_H$ is Hund's rule exchange.
Including up to second nearest neighbors we can define the
inter-atomic exchange interaction $J'(q) \equiv J'(0)-(qa)^2J'_{eff}$ where 
$a$ is the lattice constant. We also define $\bar{U}=N(0)U/3\,$,  
$\bar{J}_H=N(0)J_H/3\,$, $\bar{J}'(q)=N(0)J'(q)/3\,$, 
$\bar{U}_{eff}=\bar{U} + 2\bar{J}_H + 3\bar{J}'(0)\,$, and
$\bar{J}'_{eff}=N(0)J'_{eff}\,$ \cite{exchange}. Now 
$\chi(0)=2\mu^2_B \,S\,u_0$, where $S=1/(1-\bar{U}_{eff}\,u_0)\,$ and
$u_0=u(0)\,$.

To define the spin correlation range, $\sigma$, we first expand
Eq. (\ref{chi1}) for small q . With $u(q)\approx u_0 + u_2 (qa)^2\,$, 
Eq. (\ref{chi1}) becomes
\begin{equation}
\chi(q)=\frac{2\mu^2_B S u_0}{1 + \sigma^2 q^2} \, ,
\label{chi2}
\end{equation}
which yields a factor $\frac{1}{r}e^{-r/\sigma}$ where
$\sigma^2/a^2=S\left[u_0\bar{J}'_{eff} - u_2 \bar{U}_{eff}
\right]-u_2/u_0\,$,
in agreement with Clogston \cite{Clogston}. 

For arbitrary q we model the suceptibility with
\begin{equation}
\chi(q)=\frac{2\mu^2_B \, \chi_0(q)}
{1-I(q)\, \chi_0(q)} \, ,
\label{chi3}
\end{equation}
where $I(q)$ is a two parameter phenomenological interaction which
is determined so that Eq.(\ref{chi1}) reduces to Eq.(\ref{chi2}) for
small $q\,$. This yields
$\bar{I}(q)=N(0)I(q)=\bar{U}_{eff}/\left[1+(qa)^2 (\bar{J}'_{eff}/\bar{U}_{eff})
\right] \,$. $\bar{U}_{eff}$ is determined directly by $S$. We take
$S=3.8$ for Pt \cite{OKAnd} and find $\bar{U}_{eff}=0.737$. We fix 
$\bar{J}'_{eff}$ to provide a reasonable value for the 
spin fluctuation induced effective mass enhancement
$\lambda_{SF}$ . As in the case of Pd, the problem here is to divide the
effective mass enhancement $m^{\ast}/m=1+\lambda_{SF}+\lambda_{Ph}$
between the phonon and spin fluctuation contributions. We assume that 
$\lambda_{Ph}$ is about
the same in Pt as in Pd and take the Pd value of 
$\lambda^{Pd}_{Ph}=0.41$ .
Assuming $m^{\ast}/m-1=0.63$ for Pt \cite{OKAnd} we have 
$\lambda^{Pt}_{SF}=0.22$. Employing the standard calculation of
$\lambda_{SF}$\cite{FayAppel}, 
\begin{equation}
\lambda_{SF}=\frac{3}{2} \int_{0}^{2k_F}\,\frac{q\,dq}{2k_F^2}
\frac{[\bar{I}(q)]^2 u(q)}{1-\bar{I}(q)u(q)}\, ,
\label{lamsf}
\end{equation}
we find $\bar{J}'_{eff}=0.163$ which yields
$\sigma=3.21 \mbox{\AA}\,$. Physically, $\bar{J}'_{eff}$ is a measure of 
the range of $\bar{I}(r)$ in position space. Increasing $\bar{J}'_{eff}$ 
increases $\sigma$ and the range of $\bar{I}(r)$ but decreases the range 
of $\chi(q)$ in q-space yielding a smaller $\lambda_{SF}$ from 
Eq. (\ref{lamsf}).

 Assuming the RPA form of Eq. (\ref{chi1}) is not 
changed by the presence of scattering centers, one need only consider 
the effect of scattering on $\chi_0\,$. This was first done by de Gennes  
\cite{deGennes} for ordinary scattering alone. He showed that  $\chi_0(q)$ 
is not affected for $q=0\,$.  Fulde and Luther (FL1) \cite{FL1}
calculated $\chi(q,\omega)$  for small $q$  and Jullien \cite{Jullien}
extended this work to arbitrary $q$ and $\omega$.  Spin-orbit
scattering was later added to the ordinary scattering in FL2 \cite{FL2}.
We use the result of FL2 for the effect of s-o scattering on $\chi_0$. As 
FL2 were primarily interested in obtaining approximate analytic 
expressions for the $\omega$-dependent susceptibility we employ 
however the formalism of Julien \cite{Jullien} which is more suitable 
for computations. Equation (9) of Jullien for $\chi_0$ with ordinary 
scattering alone, can be generalized to include s-o scattering by 
comparison with Eq.(14) of FL2. The result is
\begin{equation}
\chi_0({\bf q},\omega_0)=\frac{i}{2\,\pi} \int_{0}^{\infty}\,d\omega\,
\frac{Z(\omega) \left[ 1-\pi k_F \gamma_1 Z(\omega)/3\right] }
{1 - \pi k_F \gamma_0 Z(\omega)
\left[1-\pi k_F \gamma_1 Z(\omega)/3\right] \,} ,
\label{chi0}
\end{equation}
where $Z(\omega)= \int \,\frac{d^3k}{(2\,\pi)^3}\,
G({\bf k},\omega)\,G({\bf k}+{\bf q},\omega + \omega_0 \,)$,
$\gamma_0 = 1/k_F \ell_0\,$ and $\gamma_1 = 1/k_F \ell_1\,$ with 
$\ell_0$ and $\ell_1$ the mean free paths for ordinary and s-o scattering,
respectively. The single particle propagator $G$ contains the scattering
rates  in the combination $\gamma = \gamma_0 + \gamma_1$.
We set $\omega_0 = 0$ and consider from now on only $\chi(q)$.
In their small-q approximation FL2 set $\chi_0(q)/N(0) = 1\,$. By doing
this they neglected the $\gamma_1$ corrections to $\chi_0(0)$ 
that are crucial in the following considerations. The computation of $\chi$
proceeds as in Ref.~\onlinecite{Jullien} leading to the results shown in
Fig. (1) where we take for Pt, 
$k_F=0.642\,\mbox{ cm}^{-1}$ and $a=3.923\AA\,$. Fig. (1a) shows the large 
effect of s-o scattering on $\chi_0(0)$. In Fig. (1b) $S$, $u_0$, and $\sigma$ 
are shown vs  $\gamma_1$.  $S$ and $u_0$ do not depend on  $\gamma_0$
and the dependence of $\sigma$ on $\gamma_0$ arises only through $u_2$
and is negligible.

We need the Fourier transform of $\chi(r)$ at small r to calculate the induced
spin polarization around an impurity moment and at large r  for the RKKY 
interaction responsible for spin glass transition. For 
$\chi(q) \rightarrow \chi_0(q)$ the Fourier transform can be done
analytically yielding the usual RKKY oscillations. Integrating $\chi(r)$ over 
$d^3r$ leads directly to the sum rule
\begin{equation}
\int\,d^3r\,\chi(r)=\chi(q=0)=2\mu^2_B N(0)Su_0\,.
\label{sumrule}
\end{equation}

     $\chi_0$ alone does not provide a reasonable induced moment since the RKKY
oscillations do not yield the necessary short range ferromagnetic correlations. 
This problem does not occur in our two parameter model for $\chi$ as can be
seen from Fig. (2a). Here we plot the dimensionless susceptibility 
$\bar{\chi}(r)$ which is defined by
$\chi(r)=2\mu^2_B N(0) (\Omega/a^3)\, \bar{\chi}(r)\,$, where $\Omega$ is the 
atomic volume ($a^3/4$ for the fcc lattice). The first effect of $U_{eff}$ is to 
shift the curve to larger r increasing the spin correlation range $\sigma$, an
effect discussed by Giovannini, Peter, and Schrieffer \cite{Peter}. Further 
increasing $U_{eff}$ pushes the curve above the axis for small r. 
Increasing $\bar{J}'_{eff}$ has a similar effect. The solid curve for the Pt
parameters provides both the ferromagnetic short range correlations and
the long range oscillations necessary for spin glass freezing.
The effect of scattering at small r is shown in Fig. (2b). Ordinary scattering
(dash curve) tends to smooth out the oscillations with little change in the 
area under the curve consistent with the sum rule, Eq. (\ref{sumrule}). Spin 
orbit scattering (dot-dash curve) on the other hand reduces the magnitude
of $\chi(r)$.

The giant moments observed in the bulk PtFe$_x$
\cite{{HermJLTP1}} are not seen in the Pt powders \cite{HermPRL} although,
according to susceptibility measurements, the granules contain $x=(4\pm1)$ 
ppm of magnetic impurities. The giant moment consists of two parts, 
$\mu_{gm}=\mu(i) + \mu(h)\,$, where $\mu(i=\mbox{impurity})$ is the local 
moment of the 3d electrons of the Fe impurity atom and $\mu(h=\mbox{host})$
is the spin polarization of the 5d electrons of the Pt host matrix.  We assume 
that  $\mu(i)$ of Fe in Pt has approximately
the same value as in Pd and take $\mu(i) \simeq 3\,\mu_B$. Using the 
experimental susceptibility value 
\cite{HermJLTP1}, $\mu_{gm} \simeq 8\, \mu_B$, leads to 
$\mu(h) \simeq 5\,\mu_B$.  We have
\begin{equation}
\mu(h)=4\pi \int_{0}^{r_{gm}}r^2\,dr\,\sigma_s (r)\, , 
\label{muhost1}
\end{equation}
where $r_{gm}$ is the giant moment radius and $\sigma_s (r)$ is the 
isotropic spin polarization induced by the Fe moment at $ {\bf{r}} =0$ due to the
exchange interaction $V_{ex}$ between the 3d electrons of the impurity and the 
5d electrons of the Pt host,
$\sigma_s(r)=(V_{ex}/4)\, N(0)\, \mu_B\, \bar{\chi}(r)\,\,$\cite{Clogston}.
N(0) is the DOS per spin and $\mbox{eV}\cdot \mbox{cm}^3\,$. 
In order to calculate $\mu(h)$ we need the parameters $r_{gm}$ and 
$V_{ex}$. $\mu(h)$ is not particularly sensitive to $r_{gm}$ and an
upper limit can be obtained from the sum rule, Eq.(\ref{sumrule}): 
$\mu(h)\mid_{r_{gm}\rightarrow\infty}\,=
V_{ex}\,N(0)\,a^3\,S\,u_0\,\mu_B\,/4\,$ . We take 
$r_{gm}\sim2.5\,a \sim 10 \mbox{\AA}$ as in Pd and
then fix the value of the local exchange coupling $V_{ex}$ by 
requiring that Eq. (\ref{muhost1})  yield $\mu(h)=5\,\mu_B$. We find 
$V_{ex}=2.504\,\mbox{eV}$ which is somewhat large but still seems 
reasonable.
Here we have used $N(0)=0.386(m^{\ast}_{b}/m)\,$ states/eV/atom with 
band mass $m^{\ast}_{b} /m=3.36$\cite{OKAnd}. The effect of s-o 
scattering on  
$\mu(h)$ is shown in Fig. (3), where $\mu(h)$ from 
Eq. (\ref{muhost1}) (solid curve) and for $r_{gm}\rightarrow\infty$
(dash curve) are shown versus the s-o scattering parameter 
$\gamma_1$. It turns out that $\mu(h)$ is practically independent of 
ordinary scattering, $\gamma_0$. This can be seen from the sum rule
result, $r_{gm}\rightarrow\infty$, since $S$ and $u_0$ are only affected
 by s-o scattering. Due to the rapid decrease of $\chi(r)$ in the presence
of scattering the sum rule is approximately exhausted for the
experimental $r_{gm}$. The decrease of $\chi(r)$ at small 
r seen in Fig. (2b) leads to a reduction of $\mu(h)$ by a factor 2 for 
$\gamma_1 \simeq 0.2$ and can explain why giant moments are not
observed in the Pt granules.

The spin glass freezing temperature $T_f$ in 
dilute impurity systems is determined by the long range spin polarization 
that provides the RKKY coupling between two magnetic impurities. At large 
$r$ and in the absence of scattering, $\chi(r)$ and $\chi_0(r)$ are nearly
the same and proportional to $\cos(2\,k_F\,r)/r^3$. The scale of $T_f$ is set by
the average RKKY coupling energy of a typical impurity atom pair. Although a
correct calculation of $T_f$ requires evaluation of the second moment of the 
distribution of the couplings, an estimate can be obtained from the envelope of
$\chi(r)$ determined by the peaks of the oscillations. Denoting this quantity by
$<\bar{\chi}(r_{avg})>$ we take for Fe impurities in Pt
\begin{equation}
k_B\,T_f \approx \mu^2_{\mbox{Fe}} \left( \frac{ V_{ex}}{2\mu_B} \right)^2\,2N(0)\,
\frac{\Omega}{a^3}<\bar{\chi}(r_{avg})>\,,
\label{Tf}
\end{equation}
where $\mu_{\mbox{Fe}}$ is the bare Fe moment. Without scattering,
$k_B\,T_f = \mu^2_{\mbox{Fe}}(V_{ex}/2\mu_B)^2\,2N(0)x/4\pi\,$, where 
$x=n_{\mbox{Fe}}\,/\,n_{\mbox{Pt}}$ with $n_{\mbox{Fe}}=1/r^3_{avg}$ 
and $n_{\mbox{Pt}}=4/a^3$. Using this equation with $x\approx5\,$ppm
and  $\mu_{\mbox{Fe}}=3\mu_B$, we obtain for bulk Pt a value for $T_f\,$ 
 ($2.1\,\mbox{mK/ppm}$) that is almost an order of magnitude
greater than the observed $0.26\,\mbox{mK/ppm}$.  Our rather large 
value of $V_{ex}$ presumably contributes to this discrepency. Here
however,  we are concerned with the effect of scattering on 
$T_f\,$, Eq. (\ref{Tf}). In the presence of either ordinary or s-o scattering, 
$\bar{\chi}(r)$  falls off rapidly at large r. A rough numerical fit gives
$\bar{\chi}(r) \sim \exp(-5\gamma_i\,r/a)$ for $r/a\,>\,1/\gamma_i$  where 
$\gamma_i=\gamma_0\mbox{ or }\gamma_1$. Although a power law cannot
be ruled out, the decrease is in any case much faster than $1/r^3$. 
We can thus conclude that the contributions to $\chi(r)$ we have calculated
do not lead to  a measurable $T_f$ in the presence of scattering in granular
Pt. However, at large r, diffusion-type diagrams for $\chi$ may be
dominant leading to a contribution proportional to $1/r^3$ and independent 
of ordinary scattering. In Ref.~\onlinecite{Abrahams} it was shown that these
contributions are exponentially small in the presence of s-o scattering. Thus
it seems quite reasonable that no spin glass transition was observed in the 
Pt granules.

The single grain superconducting transition temperature $T_c$ is affected by
scattering only through  the indirect effect on the spin fluctuation part pair 
interaction, $\lambda_{SF}$.  There is no direct effect for ordinary scattering 
due to Anderson's theorem which also holds for s-o scattering 
\cite{AppelOver}, for other scattering processes that obey 
time-reversal symmetry, and in zero magnetic field. In order to estimate
the indirect effect we calculate $T_c$  for s-wave pairing
with the standard weak-coupling equation\cite{FayAppel,Allen}:
\begin{equation}
T_c = \Theta_D \,  \exp\left[-\,\frac{1 + \lambda_{Ph} + \lambda_{SF}}
{\lambda_{Ph} - \lambda_{SF} - \mu^{\ast}}\right]\,.
\label{tc}
\end{equation}
Here $\lambda_{SF}$ is given by  Eq. (\ref{lamsf}) but now with scattering
included. We take $\Theta_D(Pt) = 234\, K$ and $\mu^{\ast}=0.1$ which is 
a standard estimate. We assume $\lambda_{Ph}\approx0.41$
is not affected by scattering and it turns out that $\lambda_{SF}$ is 
practically independent of $\gamma_0\,$, yielding a
decrease of only a few percent in $T_c$.  In Fig. (3) we plot $T_c$ versus
$\gamma_1$ for $\gamma_0 = 0.01$. It is seen that the $T_c$'s observed 
in Pt powders \cite{HermPRL} are reached for a s-o scattering rate 
$\gamma_1$ less than 0.1 and that $T_c$  increases strongly with 
increasing $\gamma_1\,$.

In conclusion, we have shown that ordinary and s-o scattering affect the
exchange enhanced magnetic properties of the itinerant d electrons so 
that $\mu_{gm}$ and $T_f$ of Pt$\mbox{Fe}_x\,$, e.g.,  are reduced.
On the other hand, s-o scattering weakens the spin fluctuations
to the extent that the phonons dominate and superconductivity with 
$T_c\approx 1$ mK can occur in single Pt granules. This is possible with
moderate s-o scattering since the effective electron-electron interaction in
bulk Pt is very close to zero \cite{Hauser}. Of the effects not considered 
here that could change $T_c\,$, phonon softening is probably 
the most important. To control surface phonon effects and to complement
the studies of grains an experimental search for superconductivity
in thin films of Pt is of interest. In films one must distinguish between 
thick ($\,>100\AA$) films where s-o scattering accounts for the surface
effects on the 3D electron states and thin films with smooth surfaces where
the Rashba s-o splitting occurs throughout the film thickness. How the 
spin fluctuations in very thin films of nearly ferromagnetic metals affect 
(spoil?) the Rashba effect is an open question. It would also be 
interesting to investigate thin films where the surface roughness suppresses 
the Rashba effect and s-o scattering reduces $\lambda_{SF}$.
Finally, the films can become so thin, or the grains so small, that size 
quantization of electron states occurs that may lead to new effects in the 
interplay between magnetism and superconductivity.

\acknowledgements
 
We would like to thank P. Hertel for helpful discussions. \\
\hrulefill\\
\newpage
\begin{figure}
\centerline{\psfig{file=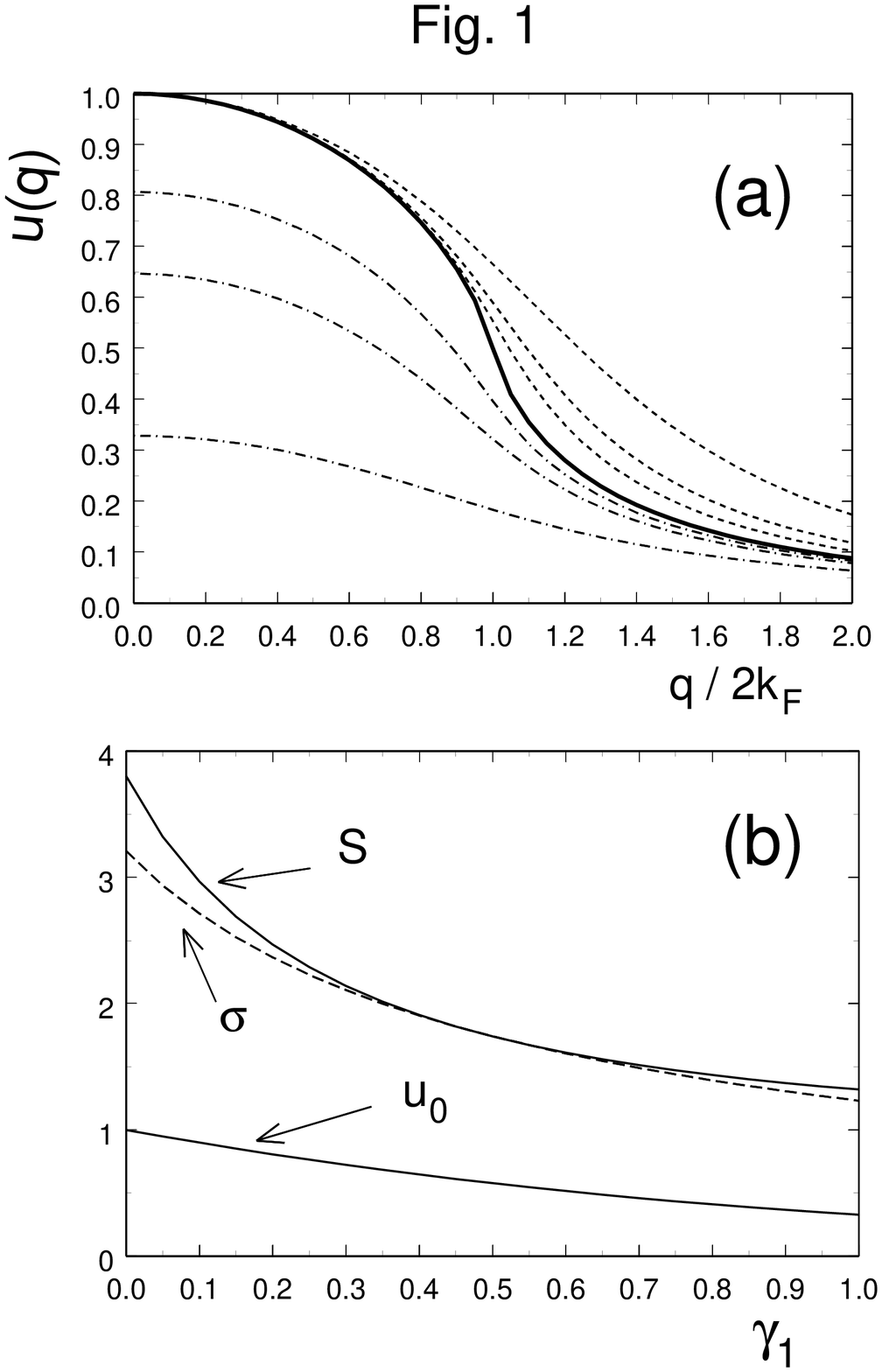,width=18cm,angle=0}}
\vskip -0.8cm
\caption{ (a) Solid curve: Lindhard function $u(q)$ without scattering. Dashed
curves: Ordinary scattering alone, $\gamma_1 = 0$, 
$\gamma_0 = 0.2, 0.4,1.0$. Dot-dash curves: s-o scattering alone, 
$\gamma_0 = 0$, $\gamma_1 = 0.2, 0.4, 1.0$. (b) Stoner factor, spin
correlation range, and $u_0 = u(0)$ as a function of s-o scattering.}
\label{fig1}
\end{figure}
\begin{figure}
\centerline{\psfig{file=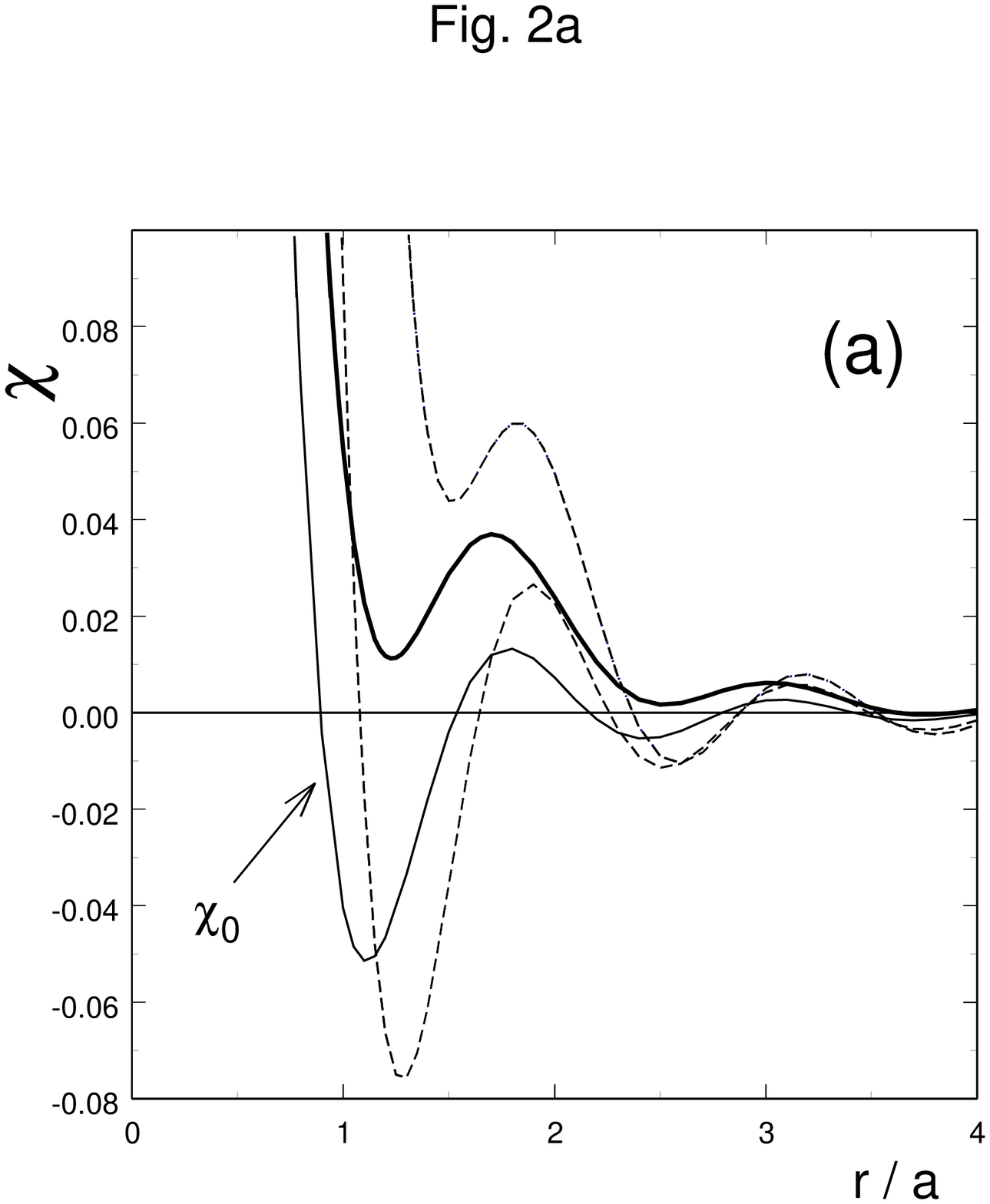,width=18cm,angle=0}}
\centerline{\psfig{file=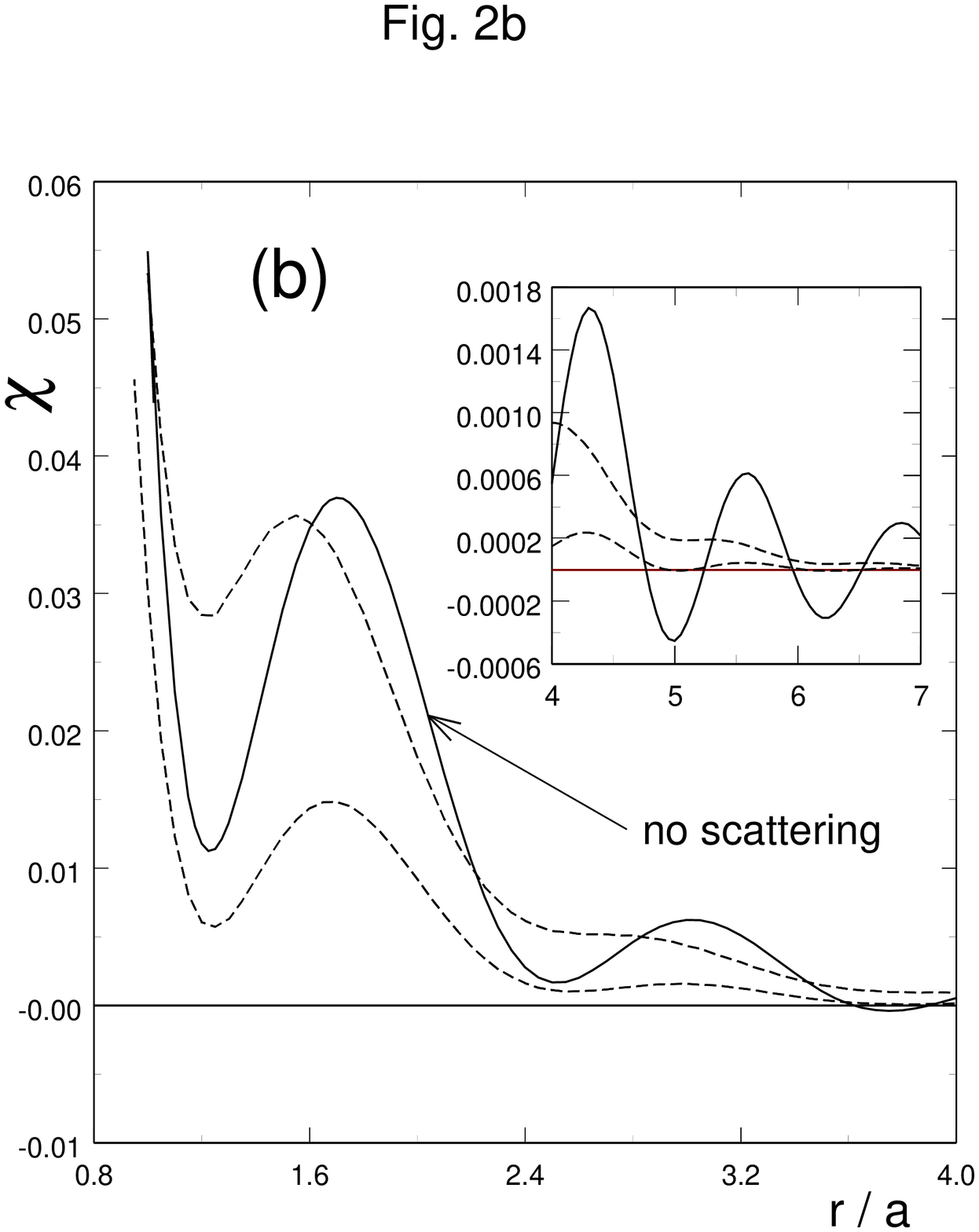,width=18cm,angle=0}}
\vskip -1.5cm
\caption{Dimensionless susceptibility $\bar{\chi}(r)$ without scattering 
vs. $r/a$. (a) Lower dash curve: $\bar{U}_{eff}=0.737, \bar{J}'_{eff}=0$; 
upper dash curve: $\bar{U}_{eff}=0.92, \bar{J}'_{eff}=0$; 
thick solid curve: Pt parameters, 
$\bar{U}_{eff}=0.737, \bar{J}'_{eff}=0.163$.  (b) $\bar{\chi}(r)$ with 
and without scattering for Pt parameters. Solid curve: no scattering; 
upper dash curve: ordinary scattering, $\gamma_0=0.2,  \gamma_1=0$; 
lower dash curve: s-o scattering, $\gamma_0=0,  \gamma_1=0.2$.}
\label{fig2}
\end{figure}
\begin{figure}
\centerline{\psfig{file=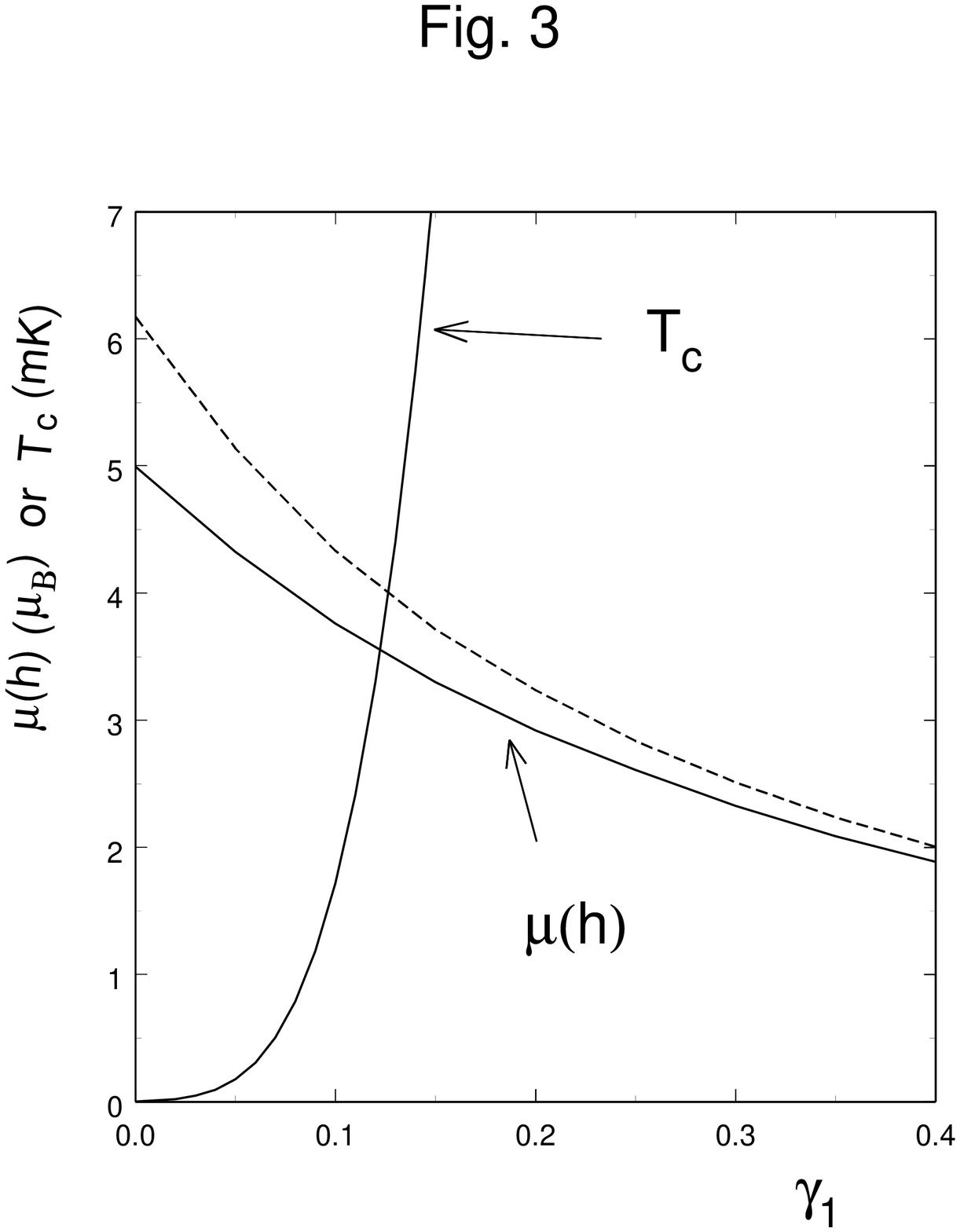,width=18cm,angle=0}}
\vskip -1cm
\caption{Superconducting transition temperature $T_c$ and the Pt host
contribution to the giant moment $\mu(h)$ as functions of the s-o 
scattering rate $\gamma_1$. The dashed curve is the exact  rule
result for infinite giant moment radius.}
\label{fig3}
\end{figure}

\begin{references}
\bibitem{exchange}For the fcc lattice and up to 2nd n.n., we have 
$J'_{eff}=J'_1+3J'_2$, where  $J'_1$ and $J'_2$ are the interatomic exchange
integrals for 1st and 2nd n.n., respectively.
\end{references}
\end{document}